\documentclass[twocolumn,prl,aps,floats,nofootinbib,showpacs]{revtex4}
\usepackage{epsfig}

\newcommand{\be}{\begin{equation}}
\newcommand{\ee}{\end{equation}}
\def\alt{\raise0.3ex\hbox{$\;<$\kern-0.75em\raise-1.1ex\hbox{$\sim\;$}}}
\def\agt{\raise0.3ex\hbox{$\;>$\kern-0.75em\raise-1.1ex\hbox{$\sim\;$}}}

\begin{document}

\title{Cosmological neutrino mass detection: The best probe of neutrino lifetime}

\author{Pasquale D. Serpico}
\affiliation{Center for Particle Astrophysics, Fermi National
Accelerator Laboratory, Batavia, IL 60510-0500, USA}

\begin{abstract}
Future cosmological data may be sensitive to the effects of a finite
sum of neutrino masses even as small as $\sim$0.06 eV, the lower
limit guaranteed by neutrino oscillation experiments. We show that a
cosmological detection of neutrino mass at that level would improve
by many orders of magnitude the existing limits on neutrino
lifetime, and as a consequence on neutrino secret interactions with
(quasi-)massless particles as in majoron models. On the other hand,
neutrino decay may provide a way-out to explain a discrepancy
$\alt0.1\,$eV between cosmic neutrino bounds and Lab data.
\end{abstract}

\pacs{13.35.Hb, %Decays of heavy neutrinos
98.80.-k,    %Cosmology
95.30.Cq    %Elementary particle processes in Fundamental astronomy and astrophysics
\hfill FERMILAB-PUB-07-015-A}

\date{23 April 2007}
 \maketitle
%%%%%%%%%%%%%%%%%%%%%%%%%%%%%%%%%%%%%%%%
{\bf Introduction }
%%%%%%%%%%%%%%%%%%%%%%%%%%%%%%%%%%%%%%%%
Recent years have seen an impressive improvement on the cosmological
constraints to the sum of neutrino masses $\Sigma=\sum m_\nu$ (for
reviews see~\cite{Lesgourgues:2006nd,Hannestad:2006zg}), with
current limits typically ranging below 1 eV, and the most aggressive
bounds (but also the most fragile ones with respect to unaccounted
systematics) already at $\Sigma\alt$0.2 eV, 95\%
C.L.~\cite{Seljak:2006bg,Fogli:2006yq}. Several forecast analyses
suggest that cosmological probes will reach in the future an
incredible sensitivity to the effects of even a tiny mass of the
cosmic background neutrinos. In particular, cosmic microwave
background (CMB) lensing extraction may be sensitive to
$\Sigma\simeq 0.035\,$eV \cite{Lesgourgues:2005yv}; CMB plus weak
galaxy lensing with tomography may also push the sensitivity to
$\Sigma$ below the level of $\sim 0.05\,$eV
\cite{Song:2003gg,Hannestad:2006as}, with an error as low as $\sim
0.013\,$eV \cite{Song:2003gg}. Also galaxy cluster surveys may probe
$\Sigma$ down to $\sim0.03\,$eV \cite{Wang:2005vr}, and a
sensitivity down to $0.05\pm0.015\,$eV may be reached combining CMB
with the data from the Square Kilometre Array survey of large scale
structures \cite{Abdalla:2007ut}. These forecasts show that
cosmology has a potential sensitivity to neutrino masses well below
the 0.1 eV level. Of course, the ultimate level of the systematics
to beat has yet to be reliably established. On the other hand, the
synergy between different strategies and probes may help to identify
the systematics, and also to improve over the above-mentioned
figures of merit.

The interest of these expectations relies on the fact that neutrino
oscillation data imply $\Sigma\agt 0.06\,$eV, where the minimum
$\Sigma\simeq 0.061\pm0.004\,$eV is attained for a normal hierarchy
(NH; values quoted at 2 $\sigma$, see~\cite{Fogli:2006yq}). For the
case of an inverted mass hierarchy (IH), the oscillation data imply
$\Sigma \simeq 0.1\,$eV. In the following, we shall proceed under
the assumption that cosmological observations will be able to probe
these Lab predictions. To be defined, we shall assume that neutrinos
have a hierarchical spectrum of either inverted or normal sign, as
favored by many theoretical models, including the simplest seesaw
ones. A degenerate mass scenario is phenomenologically allowed, with
the existing constraints given by $\Sigma < 6-7\,$eV if only
laboratory bounds from tritium endpoint~\cite{Kraus:2004zw} are
used, or $\Sigma< 0.2-2.0\,$eV from existing cosmological
observations, where the range depends on the datasets and priors
assumed~\cite{Lesgourgues:2006nd,Hannestad:2006zg}. The degenerate
scenario can be tested to some extent independently from
cosmological observations via future tritium endpoint
spectrum~\cite{Osipowicz:2001sq} and (if neutrinos are majorana
particles) neutrinoless double beta decay~\cite{Elliott:2002xe}
experiments. We want to remark, however, that our considerations
would apply qualitatively to a mildly degenerate mass pattern, too.

The main point of this paper is to motivate that, if a positive
cosmological mass detection is achieved as expected, one will be
able to put a remarkably strong constraint on the neutrino lifetime.
Note that previous attention has been paid to the cosmological
signatures of decaying neutrinos~\cite{Lopez:1998jt}. Yet the mass
range explored in those papers is very large compared to present
bounds, and the main signature considered was the impact on the
integrated Sachs-Wolfe effect on the CMB. In our considerations, the
bound comes from the impact that massive neutrinos have in the
background evolution of the universe, in a range of masses where
they are relativistic well after the CMB decoupling.

Bounds on neutrino lifetimes are usually quoted in terms of the
rest-frame lifetime to mass ratio $\tau/m$. Given a measurement in
the time interval $t$ using neutrinos with Lab energy $E$, the naive
bound which one can put is $\tau/m\agt t/E$. Using then the longest
timescale available, the universe lifetime $t_0\simeq H_0^{-1}$
(where $H_0$ is the Hubble constant), and the lowest energy
neutrinos, the ones of the cosmic background which are at least
partially non-relativistic, a bound of the order of ($m_{50}\equiv
m/50\,$meV)
\be \frac{\tau}{m}\agt \frac{1}{m\,H_0}\simeq
10^{19}\,m_{50}^{-1}\,{\rm s/eV}\,,\label{maxbound} \ee
is the strongest constraint attainable in principle. This is to be
compared with the strongest direct bound available at present given
by the observation of solar MeV neutrinos, of the order of $\sim
10^{-4}\,$s/eV ~\cite{Beacom:2002cb}, and the much stronger (but
model dependent) bound $\tau/m\agt 10^{11}\,$s/eV, which might
derive from the observations of diffuse supernova neutrino
background~\cite{Ando:2003ie}. Recently, a bound comparable to those
projections, $\tau/m\agt 4\times 10^{11}m_{50}^2\,$s/eV, has been
claimed to follow already from the requirement that the neutrinos
are free-streaming at the time of the photon decoupling, as deduced
by precise measurements of the CMB acoustic
peaks~\cite{Hannestad:2005ex}. Yet, the robustness of this
conclusion has been questioned in~\cite{Bell:2005dr}. We shall see
that the proposed bound based on cosmological neutrino mass
detection would be much closer to the maximal theoretical bound of
Eq.~(\ref{maxbound}), thus superseding by several orders of
magnitude the previous ones. More importantly, it is not based on a
model for secret neutrino interactions, but on the ``observation" of
neutrino survival, and it applies whatever the final state light
particles are.

%%%%%%%%%%%%%%%%%%%%%%%%%%%%%%%%%%%%%%%%
{\bf The bound}
%%%%%%%%%%%%%%%%%%%%%%%%%%%%%%%%%%%%%%%%
In order to estimate the bound on the neutrino lifetime from the
cosmological observation of the neutrino mass, let us recall first
how massive neutrinos affect cosmological observables. We shall base
the following discussion mostly on the treatment given
in~\cite{Lesgourgues:2006nd}, which we address for details and
further references. The main effect is due to the direct or indirect
impact of massive neutrinos in the background evolution (Friedmann
law) of the universe. In particular, in the matter epoch (but the
extension to a late dark energy dominated phase is
straightforward~\cite{Lesgourgues:2006nd}) and assuming stable
neutrinos the Friedmann equation writes
\be \frac{1}{a^2}\left(\frac{\dot{a}}{a}\right)^2=\frac{8\pi
G_N}{3}(\rho_{\rm m}+\rho_\nu)\simeq\frac{8\pi G_N}{3}\rho_{\rm
m}(1+f_{\nu})\,,\label{fried}\ee
where a dot represents a derivative with respect to the conformal
time $\eta$, $a$ is the scale factor of the universe, $G_N$ is
Newton's constant, $\rho_{\rm m}$ is the average cold dark matter
(CDM) plus baryon density, and $\rho_\nu$ is the neutrino energy
density; we have defined $f_{\nu}\equiv \rho_\nu/(\rho_{\rm
m}+\rho_\nu)$ and the second equality in Eq.~(\ref{fried}) holds to
first order in $f_\nu$. Keeping e.g. $\rho_{\rm m}+\rho_\nu$
constant, a non-vanishing $f_\nu$ today would change the epoch of
matter radiation equality $a_{\rm eq}$ with respect to the massless
neutrino case, with a scaling $a_{\rm eq}\propto (1-f_\nu)^{-1}$
(assuming ultrarelativistic neutrinos at the time of equality). This
is responsible for the main effects on the CMB anisotropy pattern,
in the range $\Sigma\alt 2.0\,$eV. Physically, postponing the time
of equality produces an enhancement of small-scale perturbations,
especially near the first acoustic peak, and increases slightly the
size of the sound horizon at recombination. When turning to the
growth of structures, there is also an additional effect. During
matter domination and on scales smaller than the free-streaming
scale, the neutrino perturbations do not contribute to gravitational
clustering, and neutrinos can be simply omitted from the Poisson
equation. On the other hand, they do contribute to the homogeneous
expansion through Friedmann equation. Therefore the exact
compensation between clustering and expansion holding for a pure CDM
scenario is slightly shifted: the balance is displaced in favor of
the expansion effect, and the gravitational potential decays slowly,
while the matter perturbation does not grow as fast as the scale
factor. Combining the continuity, the Euler, the Poisson and the
Friedmann equations one gets the evolution law for the perturbation
in the matter density field $\delta_{\rm m}$ at small-scales
\cite{Lesgourgues:2006nd},
\be \ddot\delta_{\rm m}+\frac{2}{\eta}\dot \delta_{\rm
m}-\frac{6}{\eta^2}(1-f_\nu)\delta_{\rm m}=0, \ee
which at the first order in the small parameter $f_\nu$ has a
growing mode solution of the kind $\delta_{\rm m}\propto
a^{1-\frac{3}{5}f_\nu}$. This is valid when $\rho_\nu$ is dominated
by the most massive, non-relativistic state(s), and $f_\nu\to
constant$. Physically, the combined effect of the shift in the time
of equality and of the reduced CDM fluctuation growth during matter
domination produces an attenuation of perturbations for modes $k >
k_{\rm nr}$, where $k_{\rm nr}$ is the minimum of the comoving
free-streaming wavenumber attained when neutrinos turn
non-relativistic, and given by \cite{Lesgourgues:2006nd}
\be k_{\rm nr}\simeq 1.5\times 10^{-3} m_{50}^{1/2}\,{\rm Mpc}^{-1}.
\ee

\begin{figure}[!htb]
%\vspace{-0.9pc}
\begin{center}
\begin{tabular}{c}
\epsfig{figure=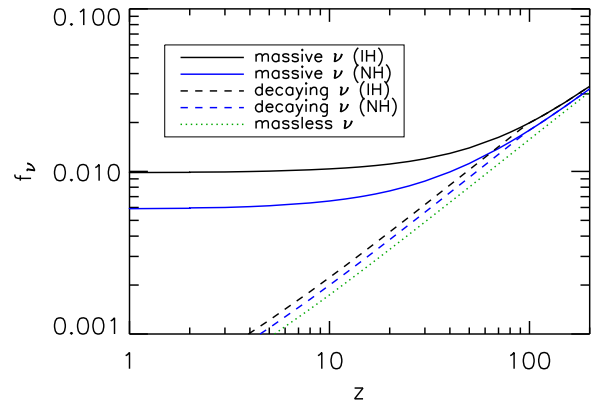,width=1.0\columnwidth}
\end{tabular}
\end{center}
\vspace{-0.9pc} \caption{The function $f_\nu(z)$ for the relevant
cosmological cases considered in the text. }\label{plot}
\end{figure}

An instantaneous decay of the massive neutrinos at an epoch $\eta_d$
in the matter era can be thought as replacing the neutrino fluid
with one having the same energy content at $\eta_d$, but whose
energy density scales from that moment on as $a^{-4}$, since the
daughter particles are relativistic. Let us estimate how large a
value of $\eta_d$, or equivalently of the proper time $t_d(=\tau$ if
the neutrino is non-relativistic), can be probed cosmologically.
Quickly after the neutrino decay one has formally $f_\nu\to 0$,
provided that $t_d\ll t_0\simeq H_0^{-1}$; from that moment on, the
cosmological effects of the decaying neutrino scenario are analogous
to the ones of a massless neutrino universe. The condition $t_d\ll
t_0$ is required by the fact that when $t_d\to t_0$, the radiation
content of the relativistic daughters of the massive neutrino has no
time to decline to zero with respect to the matter density. This
condition is necessary to change appreciably the energy budget of
the universe, thus affecting the predicted growth of the structures
and the time of equality with respect to a massive neutrino
scenario. Clearly, for a given sensitivity to the effect of neutrino
masses there is a maximum value $t_d^{\rm max}$ which would result
in a detectable change of cosmological observables. A precise
estimate of this parameter would imply a detailed forecast analysis,
which goes beyond the purpose of this paper. Yet, a simple argument
shows that, relying on the existing forecasts, a conservative lower
limit is $t_d^{\rm max}\agt t_{\rm nr}$, where $t_{\rm nr}$ is the
epoch at which the heavier neutrinos become non-relativistic, whose
redshift is defined by $m=3\,T_{\nu,0}(1+z_{\rm nr})$, $T_{\nu,0}$
being the present temperature of the neutrino gas. Indeed, when the
decay epoch satisfies $t_{\rm d}\alt t_{\rm nr}$, the energy content
of the products is the same of a relativistic neutrino fluid, and it
redshifts the same way. So, all physical effects of this scenario
are basically the same of the case where neutrinos are massless
\footnote{Cosmological probes other than big bang nucleosynthesis
(BBN) are basically insensitive to the energy spectrum of the
neutrino fluid: They are only sensitive to its overall energy
density and equation of state. In the case at hand, one or more of
the daughter particles may have a finite but much smaller mass than
the parent one, but this does not change our conclusions, at least
at our level of approximation.}. In Fig.~\ref{plot}, from top to
bottom as seen from the left side of the plot, we show $f_\nu(z)$
for the following cases: (i) a massive neutrino cosmology, where we
assume an IH neutrino mass pattern and the lightest neutrino is
massless; (ii) as in (i), but for NH; (iii) a decaying neutrino
cosmology, where massive neutrinos have IH; (iv) as in (iii), but
for NH; (v) a massless neutrino cosmology.  For the decaying cases,
we assume that all massive neutrinos decay at $t_d=t_{\rm nr}$,
where $t_{\rm nr}$ is the time of non-relativistic transition of the
heaviest neutrino state ($m\simeq 0.05\,$eV) . The neutrino mass and
mixing parameters are from \cite{Fogli:2006yq}, the cosmic neutrino
distributions are from \cite{Mangano:2005cc}, and for simplicity we
have assumed a matter-dominated cosmology with the matter density
parameter $\Omega_{\rm m}=0.24$ and the reduced Hubble constant
$h=0.73$ \cite{Yao:2006px}.

Clearly, the cases (iii), (iv), and (v) are very similar (exactly
degenerate if $t_d\ll t_{\rm nr}$) and, as long as $t_{\rm d}\alt
t_{\rm nr}$, if the massless neutrino case can be disproved, the
decaying neutrino bound immediately follows. The improvement in the
bound on the neutrino lifetime is tremendous. In particular,
neutrinos turn non-relativistic at $z_{\rm nr}\simeq
m/3\,T_{\nu,0}\simeq 100\, m_{50}$, i.e. when the universe has about
$(100\,m_{50})^{-3/2}\sim 10^{-3}m_{50}^{-3/2}$ of its present age,
and the bound is about $10^{-3}m_{50}^{-3/2}$ of the maximum
attainable limit reported in Eq.~(\ref{maxbound}),
\be \frac{\tau}{m}\agt 10^{16\,}m_{50}^{-5/2}\,{\rm
s/eV}\,.\label{lowerbound} \ee
Obviously, the previous argument does not exclude that an accurate
forecast analysis may reveal a sensitivity to a somewhat larger
$t_d^{\rm max}$. Note also that we do not require that cosmological
data need to distinguish between NH and IH: if future observations
will suggest e.g. $\Sigma=0.08\,$eV with a 1$\sigma$ error of 0.02
eV, the two neutrino mass patterns would be both consistent within 1
$\sigma$ with the best fit, yet a complete decay of neutrinos into
relativistic particles with lifetime lower than the value reported
in Eq.~(\ref{lowerbound}) could be excluded at $4\,\sigma$. Of
course, for a given cosmological sensitivity, the significance of
the above bound increases if the inverted hierarchy is realized in
nature: in that case $\Sigma\simeq 0.1\,$eV holds, and the
cosmological effects of neutrino masses are larger. Note that
accelerator neutrino experiments \cite{Barger:2000cp}, magnetized
detectors of atmospheric neutrinos \cite{Petcov:2005rv}, direct mass
searches \cite{Pascoli:2005zb}, and the serendipitous observation of
neutrinos from a galactic supernova \cite{Dighe:1999bi} may all be
used to determine the mass hierarchy. It is thus possible that by
the time cosmology will be sensitive to $\Sigma\alt 0.1\,$eV, the
hierarchy information may be available independently.

To appreciate how strong the bound of Eq.~(\ref{lowerbound}) would
be, let us consider a model of a ``secret" neutrino interaction with
a (quasi-)massless majoron field $\phi$ of the kind ${\cal
L}=g\,\bar\nu_i\nu_j\phi+\,$h.c, $i,j$ labeling different mass
eigenstates. The total decay rate for a hierarchical neutrino mass
pattern and summing over neutrino and antineutrino final state
channels is~\cite{Beacom:2002cb,Hannestad:2005ex}
\be \Gamma_{\rm d}=t_d^{-1}=\frac{g^2}{16\pi}m\, . \ee
This holds in the neutrino rest frame, but in our case this is also
the Lab decay width, give or take a factor ${\cal O}$(1), since the
neutrino is just turning non-relativistic. The constraint of Eq.
(\ref{lowerbound}) leads to the stringent bound
\be g\alt 4\times 10^{-14}\,m_{50}^{1/4}. \ee
This has to be compared with traditional bounds found in the
literature in the range $g\alt 10^{-4}\div 10^{-5}$ (see e.g.
\cite{Farzan:2002wx}), which is also a typical value invoked in the
``neutrinoless universe" scenario of Ref. \cite{Beacom:2004yd}. Even
the extremely stringent bound reported in \cite{Hannestad:2005ex} is
more than two orders of magnitude weaker.

Note that the tiny couplings which may induce the decay are not
sufficient to thermalize extra degrees of freedom in the early
universe. So, this model does not predict departure from the
standard expectation for the effective number of neutrinos $N_{\rm
eff}$ \cite{Mangano:2005cc}, which can be consistently fixed in
deriving the bound. Yet, if additional exotic physics is present, a
change (typically an increase) of $N_{\rm eff}$ is possible. The
effect of a finite $\Sigma$ can be partially compensated by an
increase in $N_{\rm eff}$, which worsens the sensitivity of
cosmological probes to $\Sigma$ (see e.g. \cite{Hannestad:2003xv}).
However, the inclusion of these exotic effects in the forecasts can
safely follow the standard parameterization used in analyses of
stable neutrino scenarios.

{\bf Discussion and conclusion} Current forecast analyses suggest
that future cosmological surveys may attain the sensitivity to
detect the effects of a sum of neutrino masses as small as
$\sim$0.06 eV, the lower limit predicted by oscillation data.
Provided that the systematics can be controlled to that level, we
have discussed in this paper how such a detection would have
profound consequences for the particle physics of the neutrino
sector, besides providing a way to measure the absolute neutrino
mass scale. In particular, when taking into account the expectations
from the Lab, excluding the $\Sigma=0$ case would improve by many
orders of magnitude the existing limits on neutrino lifetime, and as
a consequence on neutrino secret interactions with (quasi-)massless
particles as in majoron models. Strictly speaking these bounds apply
to the heaviest (or the two heaviest, in IH) mass eigenstate, but
naturaleness and phase-space considerations suggest that the
lifetime of the lightest state(s) is longer, and its coupling with a
majoron field weaker, than for the heavier one(s). Also, such a
bound would be robust with respect to the coupling mediating the new
interaction (the same may not apply to the considerations of
\cite{Hannestad:2005ex}, for example). It also applies to any
possible invisible decay channel, provided that the total mass of
the final state particles is much smaller than $\Sigma$. In
particular, this bound applies to 3-$\nu$ final state decays
$\nu_i\to\bar\nu_j\nu_j\nu_k$, as well as to decays
$\nu_i\to\nu_j+\phi$ in majoron-like models. As discussed in
\cite{Hannestad:2005ex}, a consequence of such stringent bounds is
that the decay of high energy neutrinos \cite{Beacom:2002vi}, a
target for neutrino telescopes such as IceCube, can not occur. Here
the conclusion would extend to the diffuse supernova neutrino
background, too: a disagreement with astrophysical predictions could
not be attribute to neutrino decays. We think that the idea
developed here provides a beautiful example of interplay between
particle physics, cosmological and astrophysical arguments and
motivates further the efforts to fully exploit the potential of
future cosmological surveys.

Finally, it is worth speculating briefly on the possibility that,
although future observations may attain the needed sensitivity, a
value of $\Sigma$ consistent with zero is favored\footnote{Note that
the accuracy needed to detect a finite $\Sigma$ (thus improving the
bound on the neutrino lifetime) in general differs from the accuracy
with which $\Sigma\simeq 0$ can be favored: disproving neutrino
decay may be less challenging than the opposite.}. This paper
suggests that a neutrino lifetime $\tau\alt t_{\rm nr}$, as may be
due to an extremely tiny coupling of the order of $g\agt 4\times
10^{-14}$ with a majoron-like particle, might provide a possible
explanation of a discrepancy with oscillation and Lab data, at least
if this should arise at the $\Sigma\alt0.1\,$eV level (thus
insufficient e.g. to fully explain the tension between cosmological
mass bounds and the claim of detection of neutrinoless double beta
decay \cite{Klapdor-Kleingrothaus:2004wj}). Note that this
possibility has some similarity with the ``neutrinoless universe"
scenario of Ref. \cite{Beacom:2004yd}, since it offers a way-out for
a possible non-detection of neutrino mass in cosmological data, thus
re-emphasizing the complementarity of cosmological bounds and
laboratory experiments. However, differently from the latter, it
would present no departure from the standard cosmology as early as
the BBN or CMB photon decoupling epoch. This avoids completely the
constraints based e.g. on $N_{\rm eff}$ discussed in
\cite{Beacom:2004yd} as well as the more stringent arguments put
forward in \cite{Hannestad:2005ex}.\\
{}\\
%%%
{\bf Acknowledgments} The author wishes to thank S. Dodelson for
fruitful comments, and acknowledges support by the US Department of
Energy and by NASA grant NAG5-10842.

\end{document}